\documentstyle[twoside,fleqn,espcrc2,epsf]{article}

\newcommand{\eq}{\begin{equation}}
\newcommand{\en}{\end{equation}}
\newcommand{\eqa}{\begin{eqnarray}}
\newcommand{\ena}{\end{eqnarray}}

\newcommand{\AmS}{{\protect\the\textfont2
  A\kern-.1667em\lower.5ex\hbox{M}\kern-.125emS}}

\title{Spin dependent potentials from SU(2) gauge theory}

\author{\underline{G.S.\ Bali}\address{Fachbereich Physik, Bergische 
Universit\"at, 42097 Wuppertal, Germany},
K.~Schilling\address{Forschungszentrum
J\"ulich, HLRZ, 52425 J\"ulich,
Germany}$^{\!\!\rm ,a}$
and A.~Wachter$^{\rm b,a}$\thanks{Supported
by DFG grants Schi 257/1-4 and Schi 257/3-2, and EU projects
SC1*-CT91-0642 and CHRX-CT92-0051.}}
       
\begin{document}

\begin{abstract}
We present results on spin dependent potentials from lattice
simulations of $SU(2)$ gauge theory. The Coulomb like short range part
of the central potential is identified as a mixed vector-scalar
exchange while the linear long range part is pure scalar. 
\end{abstract}

\maketitle

\section{INTRODUCTION}
\noindent
Potential models
have been found to be surprisingly successful in
describing the meson spectrum, even at rather small quark
masses, $m$~\cite{review}.
To derive the semi-relativistic Hamiltonian
one usually starts from the Bethe-Salpeter amplitude
of a two-quark bound state with quark separation $r$, takes the
static solution ($m\rightarrow\infty$) of the kernel, and expands
it in inverse powers of the quark mass around this limit.
The result can be
decomposed into the five relativistic invariants (S,V,T,A and P) with $r$
dependent (unknown) coefficient functions, the spin dependent
potentials (SDPs).

So far, SDPs have been studied
in the mid eighties~\cite{michael1,michael} on the lattice.
The applicability of the
potential approach to heavy quark physics
can be tested on the lattice by
comparing
the spectrum and wave functions,
obtained from the semi-relativistic Hamiltonian
with SDPs, to NRQCD results.
In addition we aim at replacing
phenomenologically adjusted potentials by the
``correct'' QCD potentials.

The order $1/m^2$ Hamiltonian, obtained in the manner described above,
is a generalisation of the familiar Breit-Fermi Hamiltonian.
Apart from a purely kinetic term,
it reads as follows (for equal masses),
$V_{SD}=V_{0}+\left(V_{LS}+V_{SS}+V_{vd}\right)/m^2$,
with
\begin{eqnarray}
V_{LS}(r)&=&\frac{\vec L_1\vec s_{1}-\vec L_2\vec s_{2}}{r}
\left(\frac{V_0'(r)}{2}+V_1'(r)\right)\nonumber\\
&+&\frac{\vec L_1\vec s_{2}-\vec L_2\vec s_{1}}{r}
V_{2}'(r)\quad,\\
V_{SS}(r)&=&\left((\hat r\vec s_{1})(\hat r\vec s_{2})
-\frac{\vec s_{1}\vec s_{2}}{3}\right)V_{3}(r)\nonumber\\
&+&\frac{\vec s_{1}\vec s_{2}}{3}V_{4}(r)\quad,
\end{eqnarray}
where $\vec L_i=\vec r\times \vec p_i$. $V_{vd}$ are velocity
dependent corrections.

Explicit expectation values have been associated to 
the spin-orbit and spin-spin ``potentials'' $V_1', V_2'$ and $V_3, V_4$, which 
can be computed on the
lattice~\cite{eichten,gromes}.
These potentials are related to scalar ($S$), vector ($V$) and pseudo-scalar
($P$) exchange contributions in the following way:
\begin{eqnarray}
\label{acc}
V_0(r)&=&S(r)+V(r)\\\label{ccc}
V_3(r)&=&\frac{V'(r)-P'(r)}{r}-(V''(r)-P''(r))\\
\label{ecc}
V_4(r)&=&2\nabla^2V(r)+\nabla^2P(r)\quad.
\end{eqnarray}
In addition, relativistic invariance yields the Gromes relation~\cite{gromes2},
$V_0'(r)=V_2'(r)-V_1'(r)$.

\section{LATTICE TECHNIQUES}
\noindent
From the Cornell form of the central potential,
$V_0(R)=V_0+KR-e/R$, and tree
level perturbation theory one expects
\begin{equation}
V_1'=-K\,,\, V_2'=\frac{e}{R^2}\,,\, V_3=\frac{3e}{R^3}\,,\,
V_4=8\pi e\delta(R)
\end{equation}
Since all these potentials (up to $V_1'$) are predicted to be short
ranged, we have chosen small lattice spacings to be
sensitive to the physically interesting region: we operate at
the values $\beta=2.74$ and $\beta=2.96$ that correspond to
resolutions $a=.0408(2)$~fm and $a=.0183(6)$~fm, respectively. The
physical scale has been obtained from the relation $\sqrt{K}a=440$~MeV.
More than 200 statistically independent configurations of lattice volume
$32^4$ have been collected at each coupling.

In order to reduce statistical fluctuations,
temporal links are integrated analytically wherever
possible~\cite{CERN}. The SDPs have been defined in a way
analogous to the computation of local masses
to exclude corrections proportional to $1/T$.

The lattice correlation functions,
$C(T)$, to be measured are Wilson loops with
two colour field insertions (ears) within the temporal transporters,
divided by the corresponding loop without ears ($T$ denotes the
ear-ear temporal distance).
One ear excites the gluon field between the two charges while the
second returns the field into its ground state. To obtain the SDPs,
an integration over all possible interaction times (temporal
positions of the ears) has to be performed.
The minimal distance, $\Delta$, of
an ear to an ``end'' of the Wilson loop, occurring
within the integration, represents
the time the gluon field has to decay
into its ground state after creation and, therefore, governs excited
state contaminations.
The spatial transporters within the Wilson loops
are smeared~\cite{CERN} to suppress such pollutions from
the beginning, allowing us to reduce $\Delta$: plateaus in the
ear-ear correlators have been found from $\Delta=2$ onwards
(in some cases even $\Delta=1$).

In our implementation the positions of the ears in respect to the
Wilson loop ends are kept fixed at $\Delta=2$. The
interaction time is varied by increasing the temporal
Wilson loop extent.
This results in smaller statistical fluctuations, compared
to the standard definition~\cite{eichten} in which the temporal extent
of the Wilson loop
is kept fixed and the positions of the ears are varied.

Previous authors replaced the integration over $T$ by a discrete sum.
The resulting discretization error adds to other
lattice artefacts.
We take into account that $C(T)$ depends on $T$ in a
multi-exponential way and approximate the integral from $T$ to $T+1$ by
the expression $(C(T)-C(T+1))/\log(C(T)/C(T+1))$, instead.
For $V_1'$ and $V_2'$
where $C(T)$ is weighted by an additional factor $T$
an analogous interpolating formula is used. 
The $T\rightarrow\infty$ tail of the correlator is estimated from an
exponential
fit. An integration cut-off $T_{\max}$ is chosen such that the
statistical error exceeds the estimated contribution from the
asymptotic tail by a factor four.

The lattice SDPs undergo
multiplicative renormalizations
in respect to their continuum counterparts, while $V_0$ does not.
A
non-perturbative renormalization of these quantities in the manner
suggested in Ref.~\cite{michael} is performed. 
In fig.~1 the quality of this normalization is checked by comparing
the Gromes combination $V_2'-V_1'$ (in units of the string tension, $K$)
to the force (solid curve), derived from a fit to the central potential.
We find good scaling behaviour among the two data sets
($\beta=2.74$ and 2.96) and validity
of the Gromes relation outside of the region of lattice artifacts.
A more detailed description of the
improvement tricks and data analysis can be found in Ref.~\cite{tbp}. 

\begin{figure}[htb]
\begin{center}
\unitlength1cm
\begin{picture}(7,3.6)
\put(0.0,0.0){\epsfxsize=7.0cm\epsfbox{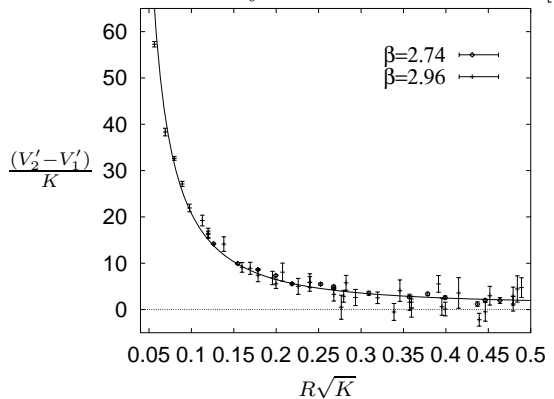}}
\put(3.5,-0.4){\mbox{\footnotesize $R\sqrt{K}$}}
\put(-0.5,2.5){\mbox{ $(V_2'-V_1')\over K$}}
\end{picture}
\end{center}
\vskip -1truecm
\caption{Test of the Gromes relation.}
\vskip -1.3truecm
\label{fig:fig1}
\end{figure}

\section{RESULTS}
\noindent
We find the central potential, $V_0$, to contain
a very clear linear confining part up to large distances
which must be of scalar
exchange type as a vector exchange can grow at most logarithmically
with $r$~\cite{gromes3}.
We confirm the second spin-orbit potential $V_2'$ to be definitely
of short range nature, leaving little room for a scalar contribution to this
potential.
The first spin-orbit potential, $V_1'$, is displayed in fig.~2.
Our lattice resolution enables us to 
establish an attractive short range contribution that can be well fitted
to a Coulomb ($1/R^2$) form, in addition to the constant long-range term,
which is in agreement with the string tension, $K$.

In principle, the potentials $V_0$, $V_3$, and $V_4$ allow for a determination
of $S$, $V$, and $P$ (eqs.~(\ref{acc})--(\ref{ecc})).
At this stage, we will assume $P$ to vanish and $V_1$
to be pure scalar (which induces the equalities $V_2=V$ and $V_1=-S$). 
This leads to a prediction for $V_3$, according to eq.~(\ref{ccc}), which
can be checked in fig.~3: we find reasonable agreement between the data sets
and the predicted curve, $V_2'/R-V_2''$, the deviations being qualitatively
understood from tree level lattice perturbation theory.
\vskip -.2truecm

\begin{figure}[htb]
\vskip -.6truecm
\begin{center}
\unitlength1cm
\begin{picture}(7,3.8)
\put(0.0,0.0){\epsfxsize=7.0cm\epsfbox{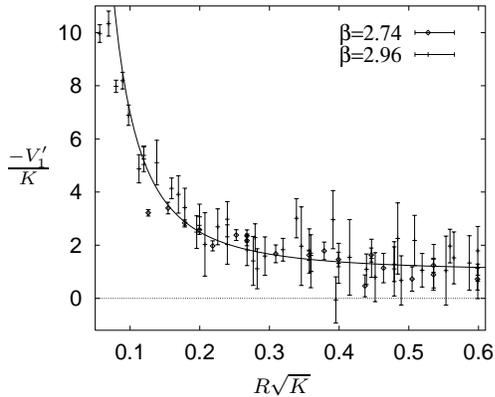}}
\put(3.5,-0.4){\mbox{\footnotesize $R\sqrt{K}$}}
\put(0.2,2.5){\mbox{$-V_1'\over K$}}
\end{picture}
\end{center}
\vskip -1truecm
\caption{$V_1'$, together with a fit curve $a/r^2+K$.}
\vskip -1.1truecm
\label{fig:fig2}
\end{figure}
\begin{figure}[htb]
\begin{center}
\unitlength1cm
\begin{picture}(7,3.8)
\put(0.0,0.0){\epsfxsize=7.0cm\epsfbox{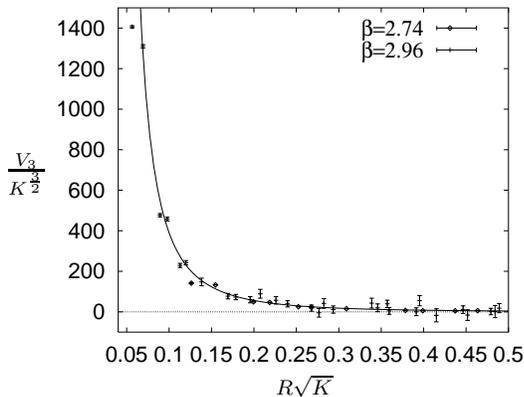}}
\put(3.5,-0.4){\mbox{\footnotesize $R\sqrt{K}$}}
\put(-0.1,2.5){\mbox{$V_3\over K^\frac{3}{2}$}}
\end{picture}
\end{center}
\vskip -1truecm
\caption{$V_3$ and $V_2'/r-V_2''$ (curve).}
\vskip -.9truecm
\label{fig:fig3}
\end{figure}

The remaining spin-spin potential, $V_4$, exhibits oscillatory behaviour as a 
lattice artifact (fig.~4) and can largely be understood as a 
$\delta$-contribution, according to $V_4=2\nabla^2V_2$.
The tree level lattice perturbative expectation $8\pi c\delta_L(\vec{R})$
is indicated by crosses in the figure. The normalization
has been obtained from a
$c/R^2$ fit to the $V_2'$ data points. The error ranges without symbols are
obtained by using single lattice gluon exchange with an infra-red
protected two-loop running coupling~\cite{tbp}. The range
corresponds to different choices of the $\Lambda$-parameter.
Apart from the dominant $\delta$-like contribution, another very short
ranged contribution seems to exist.
\vskip -.2truecm

\begin{figure}[htb]
\begin{center}
\unitlength1cm
\begin{picture}(7,4.0)
\put(0.0,0.0){\epsfxsize=7.0cm\epsfbox{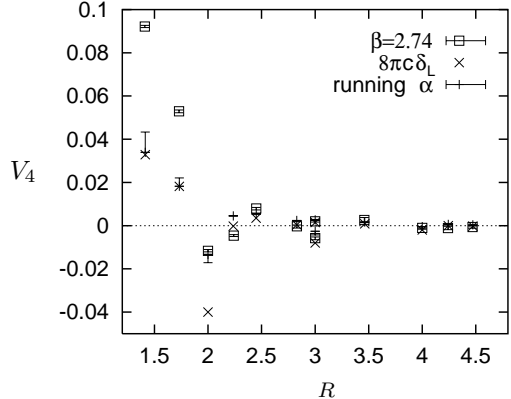}}
\put(4,-0.4){\mbox{\footnotesize $R$}}
\put(-0.1,2.5){\mbox{$V_4$}}
\end{picture}
\end{center}
\vskip -1truecm
\caption{$V_4$ in lattice units.}
\vskip -1.3truecm
\label{fig:fig4}
\end{figure}
\section{CONCLUSIONS}
\noindent
We find a consistent picture 
when identifying $-V_1$ ($V_2$) with the scalar (vector) exchange
contribution, other exchange types being negligible to order $1/m^2$.
Apart from the linear large distance part, we observe a short distance Coulomb 
like scalar contribution from the first spin-orbit potential. It appears
that the Coulomb part of the central potential splits up into a vector/scalar
ratio in between $3/1$ and $4/1$. The short range potentials can be
qualitatively understood in terms of perturbation theory.

\end{document}